\documentclass[amssymb,twocolumn,showpacs]{epl}

\usepackage{tabularx,graphicx,float}




\newcommand{\bk}{\vec{\bf k}}
\newcommand{\bq}{\vec{\bf q}}
\newcommand{\ek}{\epsilon_{\bk}}
\newcommand{\ef}{\epsilon_{\rm F}}
\newcommand{\cx}{\cos ( k_x a )}
\newcommand{\cy}{\cos ( k_y a )}

\newcommand{\susc}{\Pi ( \bq , \omega )}
\newcommand{\vf}{v_{\rm F}}

\title{Deformation of anisotropic Fermi surfaces due to electron-electron interactions.}
\author{R. Rold\'{a}n\inst{1} \and M.P. L\'{o}pez-Sancho\inst{1} \and F. Guinea\inst{1} \and S.-W. Tsai \inst{2}}

\institute{
  \inst{1} Instituto de Ciencia de Materiales de Madrid, CSIC, Cantoblanco, E-28049 Madrid, Spain.\\
  \inst{2} Department of Physics, University of California, Riverside, CA 92521, USA
}

\date{\today}

\begin{document}

\maketitle

\begin{abstract}
We analyze the deformations of the Fermi surface induced by
electron-electron interactions in anisotropic two dimensional
systems. We use perturbation theory to
treat, on the same footing, the regular and
singular regions of the Fermi surface. It is shown that, even for
weak local coupling, the self-energy presents a nontrivial
behavior showing momentum dependence and interplay with the Fermi
surface shape. Our scheme gives simple analytical expressions
based on local features of the Fermi surface.

\end{abstract}
%

%
\section{Introduction}

An open question in the study of the interactions in anisotropic
metallic systems is the deformation of the Fermi surface induced
by the interactions. The Fermi surface is one of the key features
needed to understand the physical properties of a material. Recent
improvements in  experimental resolution have led to high
precision measurements of the Fermi surface (FS), and also to the
determination of the many-body effects in the spectral function,
as reported by ARPES experiments \cite{DHS03}. The interpretation
of experiments in anisotropic strongly correlated systems remains
a complex task\cite{Zhou05}.

The FS depends on the self-energy corrections to the quasiparticle
energies, which, in turn, depend on the shape of the Fermi
surface. Hence, there is an interplay between the self-energy
corrections and the FS topology. For weak local interactions, the
leading corrections to the FS arise from second order diagrams.
The self energy, within this approximation, can show a significant
momentum dependence when the initial FS is anisotropic and lies
near hot spots (see below). This simultaneous calculation of the
FS and the second order self energy corrections is a formidable
task. Many approaches have been used to study  this problem like
pertubation theory\cite{ViRu90}, bosonization
methods\cite{AHCN94,FSLut99}, or perturbative Renormalization
Group calculations\cite{Metz03,Kop03,FCF05}, and the cellular
dynamical mean-field theory (CDMFT), an extension of Dynamical
Mean Field Theory\cite{Kotl05}.

In this work, we calculate perturbation theory corrections and use
Renormalization Group arguments\cite{S94,MCC98} in order to study
analytically the qualitative corrections to the shape of the FS
induced by the electron-electron interaction. This method allows
us to classify the different features of the FS from the
dependence of the self-energy corrections on the value of a high
energy cutoff, $\Lambda$, defined at the beginning of the
Renormalization process. As it will be shown later, one can also
analyze the effects of variations in the Fermi velocity and the
curvature of the non interacting FS on the self-energy
corrections.

We will study two dimensional Fermi surfaces, and we will consider mostly the
$t-t'$ Hubbard model, although the calculations do not depend on the
microscopic model which gives rise to a particular Fermi surface. We define
the model in the next section. Then, we describe the way the corrections
induced by different features of the
FS depend on the high energy
cutoff $\Lambda$.
Next, we present a detailed calculation of the
changes expected for a regular
FS, and make contact with results
from ARPES experiments on cuprates. At the end we highlight the
most relevant aspects of our calculation, and compare them with results
obtained using alternative schemes.

\section{The model}

\begin{figure}[h]
\twoimages[width=8cm]{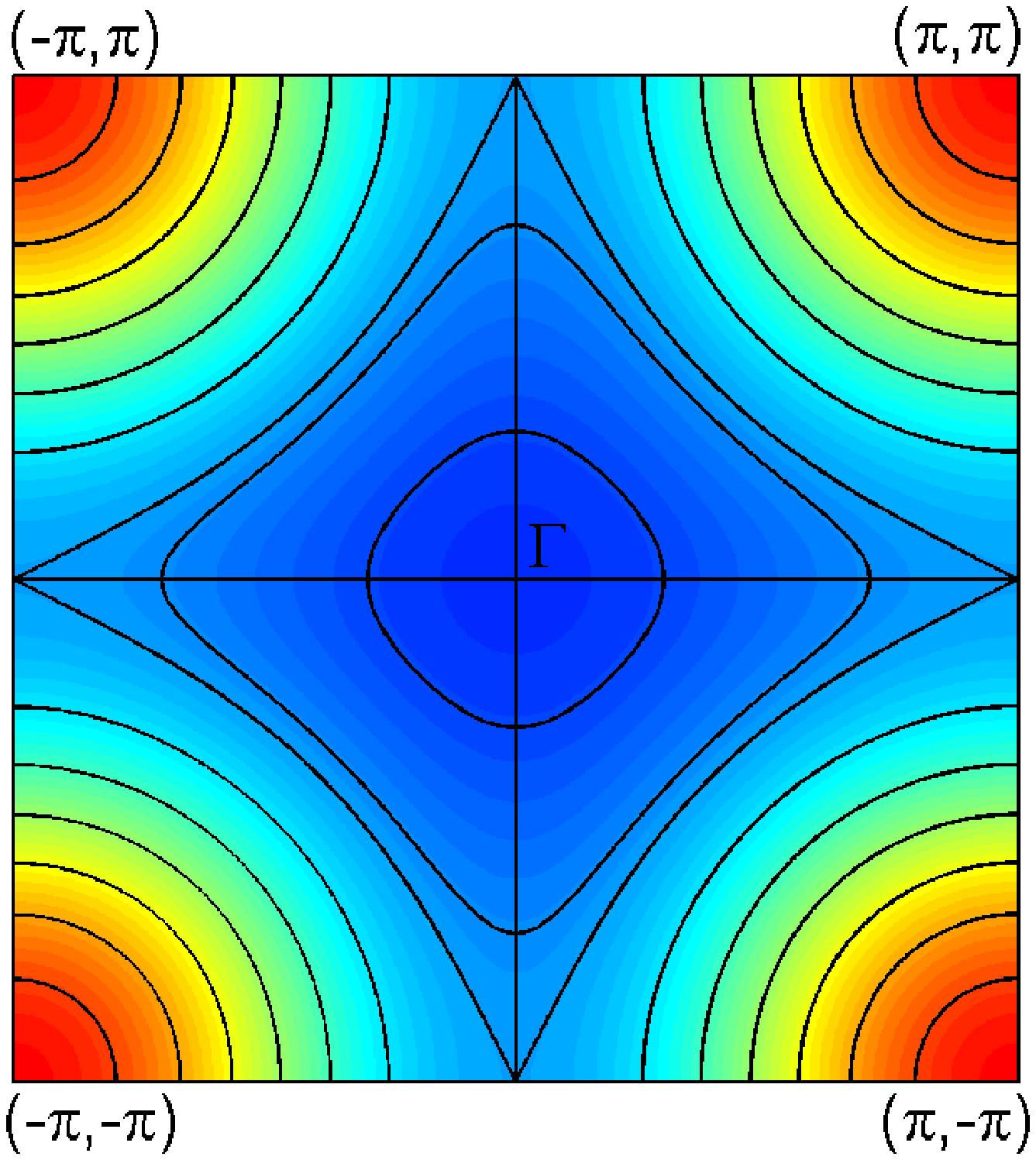}{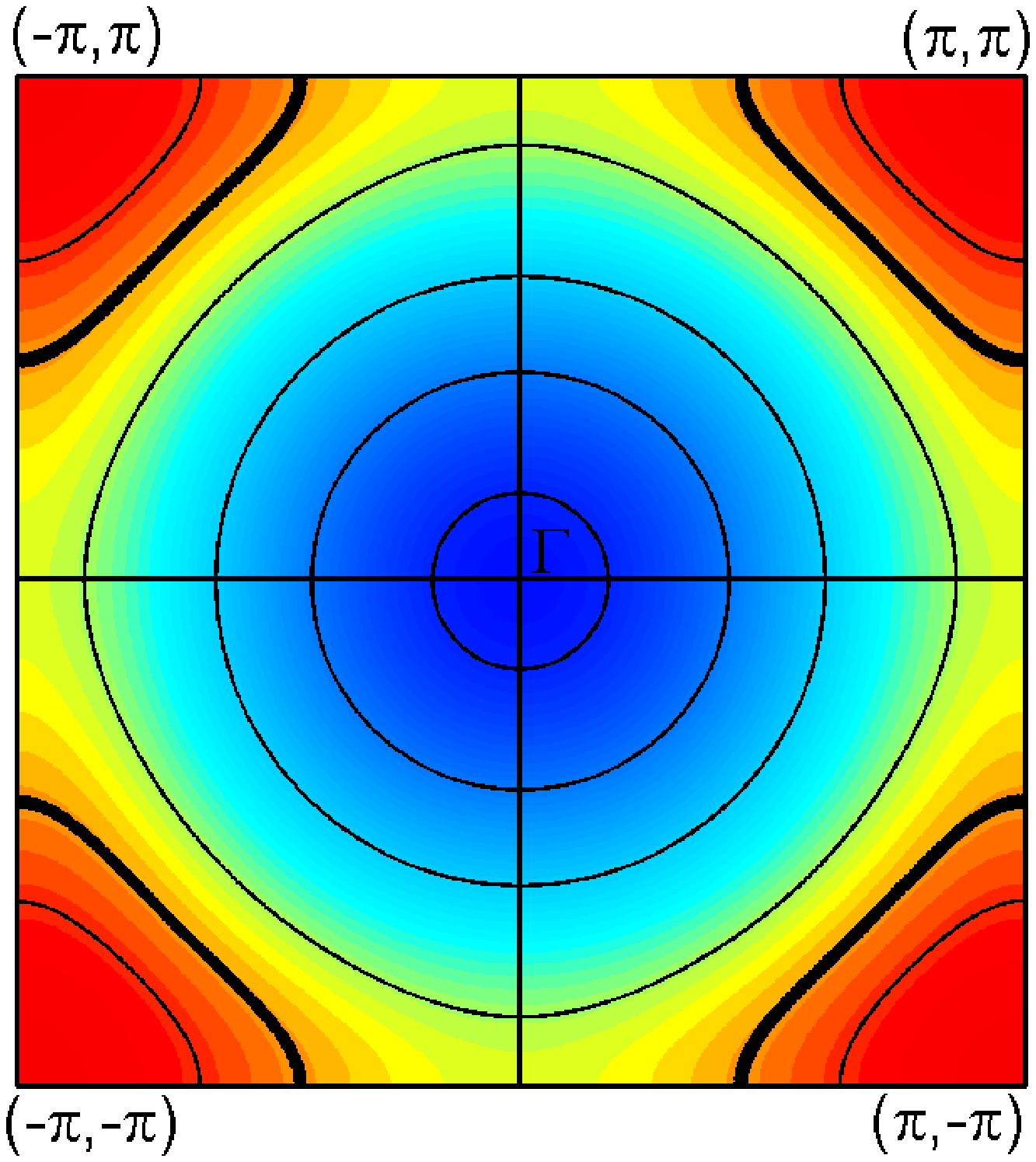}
\caption{ Qualitative picture of the evolution of the FS with
filling from almost isotropic to convex, going through a FS
exhibiting inflexion points, and one with van Hove singularities
(left panel, $t'=-0.3t$). A region with almost perfect nesting is
shown in the right panel ($t'=0.3t$).} \label{FS-Desnudas}
\end{figure}

The hamiltonian of the  $t-t'$ Hubbard model is:
\begin{equation}
{\cal H}\!=\!t\!\sum_{s; i,j \, {\rm n. n.}}\!\! c^\dag_{s,i} c_{s,j} +
t'\! \sum_{s; i,j \, {\rm n. n. n.}}\!\! c^\dag_{s,i} c_{s,j} + U\! \sum_i\!
n_{i \uparrow} n_{i \downarrow} \label{hamil} \end{equation}
where $ c_{s,i} (c^\dag_{s,i})$ are destruction (creation) operators
for electrons of spin $s$ on site $i$, $n_{i,s}=c^\dag_{s,i} c_{s,i}$ is the number
operator, $U$ is the on-site repulsion, and $t$ and $t'$ are the nearest
and next-nearest neighbors hopping amplitudes, respectively.
The Fermi surfaces of the non interacting systems are defined by:
\begin{equation}
\ef = \varepsilon (\bk) = 2 t \left[ \cx + \cy \right] + 4 t' \cx
\cy \label{dispersion} \end{equation} where $a$ is the lattice
constant (see Fig.[\ref{FS-Desnudas}]).

Assuming that $t < 0 , t' > 0$ and $| 2 t' | < | t |$, the Fermi
surface is convex for $- 2 t + 4 t' \le \ef \le \epsilon_0 = - 8
t' + 16 t'^3 / t^2$. For $ - 8 t' + 16 t'^3 / t^2 \le \ef \le - 4
t'$  the Fermi surface shows eight inflection points, which begin
at $k_x  = k_y  = k_0 = a^{-1} \cos^{-1} ( - 2 t' / t )$ and move
symmetrically around the $( \pm 1 , \pm 1 )$ directions, towards
the center of the edges of the square Brillouin zone, $( 0 , \pm
\pi ) , ( \pm \pi , 0 )$. For $\ef = 4 t'$ the Fermi surface
passes through the saddle points (van Hove singularities) located
at these special points of the Brillouin zone. For $4 t' < \ef \le
- 4 t$, the Fermi surface is convex and hole like, centered at the
corners of the Brillouin Zone, $ ( \pm \pi , \pm \pi )$ (See
Fig.[\ref{FS-Desnudas}]). Finally, when $t'=0$, the model has
particle hole symmetry, and the Fermi surface shows perfect
nesting for $\ef = 0$.

The effects of the Hubbard interaction when the Fermi surface is near perfect
nesting\cite{ZYD97,AD97,F03,F03b,FCF03} or near a Van Hove
singularity\cite{LB87,F87,D87,S87,LMP87,MG89,Netal92,GGV96} have
been extensively studied. Anomalous effects are also expected when the Fermi
surface has inflection points\cite{GGV97b,FG02}.

\section{Self-energy corrections to singular Fermi surfaces}

\begin{figure}[b]
\onefigure[width=6cm]{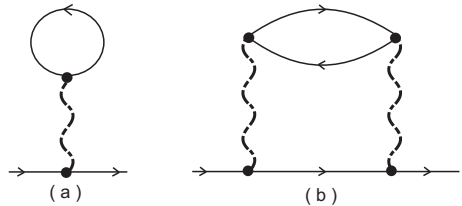} \caption{Low order self
energy diagrams. Left: Hartree diagram. Right: two loop
correction.} \label{selfenergy}
\end{figure}

The corrections to the non interacting Fermi surface are given by
the real part of the self energy. Using second order perturbation
theory, the self energy is given by the diagrams shown in
Fig.[\ref{selfenergy}]. In the following, we assume that the
effect of the high energy electron-hole pairs on the
quasiparticles near the Fermi surface have been integrated out,
leading to a renormalization of the parameters $t , t'$ and $U$ of
the hamiltonian, neglecting the possibility that other couplings
are generated. Thus, the hamiltonian, eq.(\ref{hamil}), describes
low temperature processes below a high energy cutoff, $\Lambda \ll
t , t'$. For consistency, $U \leq \Lambda$. We also assume that
the Fermi surface of the interacting system exists, and that it
has the same topology as that of the non interacting system.

The frequency dependence of the
imaginary part of the self energy in a nested region of
the Fermi surface, or at van Hove singularities
is known to be linear, unlike the usual quadratic dependence expected in
Landau's theory of a Fermi liquid.

\begin{equation} {\rm Im} \Sigma_2 ( \bk , \ek ) \propto | \ek |
\label{hotspot}
\end{equation}
where $\ek=\varepsilon (\bk)-\ef$. Away from the hot spots, the
leading contribution to the two loop self energy, when the Fermi
surface is near a van Hove singularity, comes from diagrams where
the polarizability bubble, $\susc$, involves transitions near the
saddle point\cite{HR95}. Near a nesting situation, the
polarizability at low momenta is similar to that of a one
dimensional Fermi liquid. The susceptibilities can be written as:
\begin{equation} \susc \sim \left\{ \begin{array}{lr} W^{-1}
\tilde{\Pi}_{\rm vH}
\left( \frac{\omega}{ m^* | \bq |^2} \right) &{\rm van \, \, Hove} \\
W^{-1} \tilde{\Pi}_{\rm 1D} \left( \frac{\omega}{v_{\rm F} | \bq
|} \right)&{\rm nesting}
\end{array} \right. \label{susc} \end{equation} where $v_{\rm F}$ is the
normal Fermi velocity in the nesting situation, and $m^*$ is an average of the
second derivative of the bands at the saddle point. Note that, in both cases, the density of states is
proportional to $W^{-1} \sim t^{-1} , t'^{-1}$.

 The imaginary part of the second order self energy near the regular regions
of the Fermi surface can be written as\cite{HR95}:
\begin{equation} {\rm Im} \Sigma_2 ( \bk , \ek ) \sim
\int_0^{\ek} d \omega \int_0^{q_{\rm max}} d q \, {\rm Im} \Pi ( q
, \omega ) \label{HR} \end{equation} where $q_{\rm max} \sim |
\Lambda | / v_{\rm F}$, and $v_{\rm F}$ is the Fermi velocity in
these regions.
Using eq.(\ref{susc}),  we find:
\begin{equation}
{\rm Im} \Sigma_2 ( \bk , \ek ) \propto \left\{ \begin{array}{lr}
\ek^{3/2} &{\rm van \, \, Hove} \\ \ek^2 &{\rm nesting}
\end{array} \right. \label{susc_reg} \end{equation}
We recover the
usual Fermi liquid result for the regular parts of the Fermi
surface near almost nested regions. This result arises from the
fact that the small momentum response of a quasi--one--dimensional
metal does not differ qualitatively from that predicted by
Landau's theory of a Fermi liquid.

Finally, near an inflection point, we can use the techniques developed
in\cite{GGV97b,FG02} to obtain:
\begin{equation}
{\rm Im} \Sigma_2 ( \bk , \ek ) \propto \ek^{3/2}
\end{equation}
It is finally worth noting that there is another special point, where the
Fermi surface changes from convex to concave and a pair of inflection points
are generated for $\ef = \epsilon_0$ and ${\bf \vec{k}} \equiv ( k_0 , k_0 )$
defined earlier. At this point, the imaginary part of the self energy behaves
as ${\rm Im} \Sigma_2 ( \bk , \ek ) \propto \ek^{5/4}$.

We can obtain the real part of the self energy from the imaginary
part by a Kramers-Kronig transformation, and restricting the frequency
integral to the interval $0 \le \omega \le \Lambda$. We obtain:
\begin{equation}
{\rm Re} \Sigma_2 ( \bk , \ek ) \propto - g^2 | \Lambda | \times
\left\{
\begin{array}{lr} \log^2 \left( \frac{\Lambda}{\ek} \right) &{\rm van \, \, Hove}
\\ \log \left( \frac{\Lambda}{\ek} \right) &{\rm nesting} \end{array}
\right. \label{real_hot}
\end{equation} where the negative sign is due to the fact that
it is a second order contribution in perturbation theory, and $g$
is a dimensionless coupling constant of order $U/W$. The sign is
independent of the sign of $U$ in eq.(\ref{hamil}). In the regular
parts of the Fermi surface, eq.(\ref{susc_reg}) leads to:
\begin{equation}
{\rm Re} (\bk,\ek) \propto \left\{ \begin{array}{lr} - g^2 \frac{|
\Lambda |^{3/2}}{W^{1/2}} &{\rm van \, \, Hove} \\ - g^2 \frac{|
\Lambda |^2}{W} &{\rm nesting}
\end{array} \right. \label{real_reg} \end{equation} where the
additional powers in $W$ arise from the $m^*$ and $v_{\rm F}$ factors
in the susceptibility, eq.(\ref{susc}).

In the limit $\Lambda / W \rightarrow 0$, the different dependence on
$\Lambda$ of the self-energy corrections at different regions of the Fermi
surface is enough to give a qualitative description of the changes of the
Fermi surface. For instance, when the non interacting Fermi surface is close
to the saddle point, ${\bf \vec{k}} \equiv a^{-1} ( \pm \pi , 0 ) , a^{-1} (
0 , \pm \pi )$, the self-energy correction is negative and highest in this
region. Note that the logarithmic divergences in eq.(\ref{real_hot}) are
regularized
by the temperature or elastic scattering.

In order to remove the Fermi surface from a van Hove point or
nesting situation, a large number of electrons must be added to
the regular regions. When the points of the FS near these hot
spots are at distance $k$ from the hot spot, the change in the
self energy needed to shift the Fermi momentum by an amount
$\delta k$ is, using eq.(\ref{real_hot}), $\delta \Sigma \propto
g^2 \Lambda \frac{\delta k}{k}$
with additional logarithmic corrections near a van
Hove singularity. Near the regular regions of the Fermi surface, a
shift in energy of order $\delta \Sigma$ leads to a change in the
momentum normal to the Fermi surface of magnitude
$\delta k_{\rm reg} \sim \delta \Sigma / v_{\rm F}$. The area
covered in this shift gives the number of electrons which are
added to the system near the regular regions of the Fermi surface.
We find $\delta n \sim k_{\rm max} \delta k_{\rm reg} \sim
g^2 \frac{k_{\rm max} \Lambda}{v_{\rm F}}  \frac{\delta k}{k}
\label{shift_n}$
where $k_{\rm max} \sim a^{-1}$
determines the size of the regular regions of the Fermi surface.
The value of $\delta n$ diverges as the Fermi
surface moves towards the hot spot, $k \rightarrow 0$.
Hence, the
number of electrons needed to shift the FS away from the
hot spot also diverges. This analysis essentially reproduces the
calculation at fixed chemical potential in the presence of a
reservoir with regular self-energy corrections given
in\cite{GGV96,G01} (a different analysis\cite{IKK02} does not make
use of a reservoir).

\section{Self-energy corrections to regular Fermi surfaces} We study now the system at a
filling which yields a curved FS, slightly anisotropic, in the
absence of singularities. Near the Fermi surface the electronic
dispersion can be approximated by:
\begin{equation}\label{epsilon-k}
\ek = v_{F}\kappa_{\perp}+\beta\kappa_{\parallel}^{2}
\label{dispersion_2}
\end{equation}
where $\kappa_{\parallel}$ is the momentum parallel to the FS,
$\kappa_{\parallel}=(\bk-\bk_F)_{\parallel}$, $\kappa_{\perp}$ is
the momentum perpendicular to the FS relative to $k_F$,
$\kappa_{\perp}=(\bk-\bk_F)_{\perp}$, $v_F$ is the Fermi velocity
$v_F=\hat{\mathbf{n}}_{\perp}\cdot\mathbf{\nabla}\varepsilon(\bk)$
and $\beta$ is related to the local curvature of the Fermi surface
$b=\hat{\mathbf{n}}_{\parallel} \cdot
\left(\mathbf{\nabla}^2\varepsilon(\bk)\right)
\hat{\mathbf{n}}_{\parallel}$, by $\beta=bv_F/2$. The Fermi
velocity $v_F$ and the FS curvature $b$, are functions of $t , t',
\ef$ and the position along the Fermi line. We assume that the
main contribution to the self energy arises from processes where
the momentum transfer is small, or from processes which involve
scattering from the region under consideration to the opposite
part of the Fermi surface, {\it i. e.}, backward scattering. This
assumption can be justified by noting that the Hubbard interaction
is momentum independent, so that the leading effects are
associated to the structure of the density of states. The
processes discussed here have the highest joint density of states.

Using the parametrization in eq.(\ref{dispersion_2}), the imaginary part of the
self energy, which describes the decay of quasiparticles in the region under
consideration, is independent of the cutoff $\Lambda$. The contribution from
forward scattering processes is:
\begin{equation}\label{Im-Sigma-Forw}
Im\Sigma_2(\bk,\omega)=\frac{3}{64}\frac{U^2
a^4}{\sqrt{2}\pi^2}\frac{\omega^2}{v_F^2|\beta|} \ .
\end{equation}
The quadratic dependence of energy is expected, and consistent
with Landau's theory of a Fermi liquid. This contribution diverges
as $\vf \rightarrow 0$, that is, when the Fermi surface approaches
a van Hove singularity, or as $| b | \rightarrow 0$ which signals
the presence of an inflection point or nesting. The contribution
due to backward scattering is exactly the same as that from
forward scattering, eq.(\ref{Im-Sigma-Forw}), with the same
numerical prefactors.

Using a Kramers-Kronig transformation, and integrating again in
the interval $0 \le \omega \le \Lambda$, we obtain:
\begin{equation}
Re\Sigma_2(\bk,\omega=0)=-\frac{3}{64}\frac{U^2
a^4}{\sqrt{2}\pi^3}\frac{\Lambda^2}{v_F^2|\beta|}
\end{equation}
This expression gives the leading corrections to the shape of the
FS. Notice that both, $\vf$ and $\beta$, are momentum dependent.

\begin{figure}
\onefigure[width=14cm]{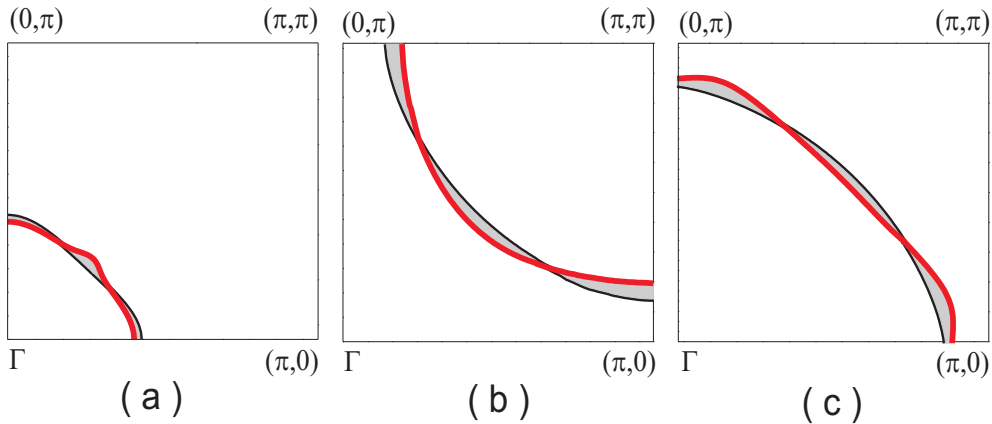}

\caption{Deformations induced by the interactions on the FS
 of the $t-t'$ Hubbard model, in the first quadrant of the
Brillouin zone.
Thin black line represents the unperturbed FS while the thick red
line represents the FS corrected by the interaction, the shadowed
region corresponds to their difference. For $t'/t=-0.3$  (a): high
doping range and (b) close to half filling. For $t'/t=+0.3$
(c)close to half filling. } \label{holeFS}
\end{figure}

In Fig.[\ref{holeFS}](a)-(b), the bare (thin black line) and
renormalized (thick red line) Fermi surfaces are shown at two
different densities in the first quadrant of the Brillouin zone
(BZ) for the parameter values $t=-1$, $t'=-0.3t$, reminiscent of
hole-doped cuprates. At the density shown in Fig.[\ref{holeFS}](a)
the self-energy corrections are stronger in the less curved
regions of the FS around the diagonal parts of the BZ. This
correction coincides with that found in \cite{FCF04} for the
renormalization of a flat FS by a two-loop field theory RG
approach, where interactions induce a small curvature to the bare
flat FS. When we change the filling, the FS shape varies, and
close to half-filling the FS has the form shown in
Fig.[\ref{holeFS}](b). The change in shape qualitatively agrees
with the doping evolution of $k_F$ measured by ARPES on cuprates
\cite{Yosh,Kami}. The self energy corrections, close to
half-filling, enhance the hole-like curvature and flatten the FS
close to the $(\pi, 0)$ and $(0, \pi)$ points of the BZ as shown
in Fig.[\ref{holeFS}](b).

In  Fig.[\ref{holeFS}](c) we show the FS corresponding to $t=-1,
t'=0.3t$ ($t/t' > 0$) reminiscent of the electron-doped cuprates,
close to half-filling, at a similar density as the one represented
in Fig.[\ref{holeFS}](b). The self-energy corrections here are
stronger at the most curved regions of the FS, in the proximity of
the saddle points (where $v_F$ diverges). The corrected FS is
closer to a nesting situation than the bare FS. Our results, near
half-filling,  are in overall agreement with those of
\cite{Kotl05}, although we find that the self energy corrections
are stronger at the antinodal region in both hole-like and
electron-like Fermi surfaces.

\section{Conclusions}

We have presented a simplified way of taking into account the
self-energy corrections to the Fermi surface. We have made use of
the different dependences of the self energy on the high energy
cutoff in order to analyze the main features of the changes of the
FS. The results suggest that the main self-energy corrections,
which are always negative, peak when the FS is close to the $( \pm
\pi,0 ) , ( 0 , \pm \pi )$ points in the Brillouin zone. If these
contributions are cast as corrections to the hopping elements of
the initial hamiltonian, we find that the nearest neighbor
hopping, $t$,  is weakly changed (as it does not contribute to the
band dispersion in these regions). The next nearest neighbor
hopping, $t'$ which shifts the bands by $- 4 t'$ in this region,
acquires a negative correction. This implies that the absolute
value of $t'$ grows when $t' > 0$, or decreases, when $t' < 0$, in
reasonable agreement with the results in\cite{Kotl05}. Note that
the tendency observed in our calculation towards the formation of
flat regions near these points, when analyzed in higher order
perturbation theory, will lead to stronger corrections. Our
results also confirm the pinning of the FS near saddle points, due
to the interactions. The analysis is consistent with  the measured
Fermi surfaces of the cuprates\cite{DHS03} and qualitatively agree
with the doping evolution reported by ARPES\cite{DHS03, Yosh,
Kami}.

Finally the analysis presented here is valid only at weak
coupling, and we do not consider corrections to the interactions
or to the wave-function renormalization. The main goal of this
scheme is that can be used with different models and the
expressions obtained are analytical and related to the local
features of the non-interacting FS in a simple way, so that they
can be readily used to get an estimate of the corrections
expected.

\acknowledgments

Funding from MCyT (Spain) through grant MAT2002-0495-C02-01 is
acknowledged. R.R. and F. G. would like to thank the hospitality
of Boston University, where part of this work was done. We
appreciate useful conversations with A. H. Castro Neto, A. Ferraz,
J. Gonz\'alez, G. Kotliar, and M.A.H.Vozmediano.

\end{document}